\documentclass[useAMS,usenatbib]{mn2e}
\usepackage{latexsym,amsmath,amssymb,enumerate}
\usepackage[]{graphicx,lscape}
\usepackage{caption}
\usepackage{subfigure}
\usepackage{multirow}

\usepackage{color,verbatim}
\usepackage{ulem}

\title[Dynamical Modeling of NGC 6809]{ Dynamical Modeling of NGC 6809: Selecting the best  model using Bayesian Inference}
\author[F. I. Diakogiannis, G. F. Lewis and R. A. Ibata]{
Foivos I. Diakogiannis$^{1}$\thanks{E-mail:
f.diakogiannis@physics.usyd.edu.au}, Geraint F. Lewis$^{1}$    and Rodrigo A. Ibata$^{2}$\\
$^{1}$Sydney
Institute for Astronomy, School of Physics, A28, The University of Sydney, NSW 2006, Australia\\
$^{2}$Observatoire astronomique de Strasbourg, Universit\'{e} de Strasbourg, CNRS, UMR 7550, 11 rue de l' Universit\'{e},\\ 
\phantom{$^{2}$}F-67000 Strasbourg, France}
\begin{document}


\pagerange{\pageref{firstpage}--\pageref{lastpage}} \pubyear{2013}

\maketitle

\label{firstpage}

\begin{abstract} 
The precise cosmological origin of globular clusters remains uncertain, a situation hampered by 
the struggle of observational approaches in conclusively identifying the presence, or not, of dark 
matter in these systems.
In this paper, we address this question through an analysis of the particular case of NGC 6809. 
While previous studies  have performed dynamical modeling of this globular cluster using a small 
number of available kinematic data, they did not perform appropriate statistical inference tests for the choice 
of best model description; such statistical inference for model selection is important since, in general, different models can 
result in significantly different inferred quantities.   
With the latest kinematic data,
we use Bayesian inference tests for model selection and thus obtain the best fitting 
models, as well as  mass and dynamic mass-to-light ratio estimates.  
For this, we introduce a new likelihood function that provides more constrained distributions for the defining parameters of dynamical models. 
Initially we consider models with a known distribution function, and then  model the cluster using solutions of the spherically symmetric Jeans  equation; 
this latter approach  depends upon the mass density profile and anisotropy $\beta$ parameter.    
In order to find the best description for the cluster we compare these models by calculating their Bayesian evidence.   
We find smaller mass and dynamic mass-to-light ratio values than previous studies,   with the best fitting Michie model for a constant mass-to-light ratio of $\Upsilon = 0.90^{+0.14}_{-0.14}$ and $M_{\text{dyn}}=6.10^{+0.51}_{-0.88} \times  10^4 M_{\odot}$.   We  exclude the significant presence of dark matter throughout the cluster, showing that  no physically motivated distribution of dark matter can be present away from the cluster core. 
\end{abstract}


\section{Introduction}

An unresolved problem of modern astrophysics is the question on the formation of star clusters. Current evidence can not conclusively identify whether globular clusters (GCs) contain any significant  amounts of dark matter (hereafter DM) as a result of their formation history. \cite{1984ApJ...277..470P} stated, based on reasonable hypotheses, that we expect to find dark matter halos in two spatial scales, namely galaxies and globular clusters, thus suggesting that GCs form in their own DM halos. Several authors argued against this (e.g. \citealt{2011ApJ...743..167B}, \citealt{2011ApJ...741...72C}), although it is quite possible that GCs were originally created in their DM halo that was subsequently stripped off by tidal forces \citep{2003MNRAS.346L..11B} leaving a DM dominated core. 

Recently, \cite{2013MNRAS.428.3648I} performed a dynamical analysis of the globular cluster NGC 2419 using two different methods, namely models based upon probability distribution functions, and a new method for solving  the spherically symmetric Jeans  equation.  Each approach had different results: one method predicted no DM component,  while the other, which allows for much greater freedom of the cluster models, permitted a small DM core. Unfortunately the complexity of the latter method did not allow for a quantitative comparison between the two models, thus leaving this an open problem. 

The purpose of the present work is two-fold: first we use the most comprehensive kinematic data  published to date (\citealt{2010MNRAS.401.2521L}; hereafter LA10) in order to perform more accurate estimates of dynamic mass and dynamic mass-to-light ratios. Second, we address the question whether NGC 6809 (also known as M55) contains any significant amount of dark matter. For this, we   model dynamically the cluster using several  methods that demonstrate different freedom in the behavior of corresponding physical quantities. First we use models based on   probability density functions (hereafter PDFs). We then model the cluster using solutions of the spherically symmetric Jeans equation. 

Using these physically motivated models we aim to robustly determine the quantity of dark matter in NGC 6809.  
For all the above mentioned methods, we propose a new functional form for the likelihood which takes into account the full kinematic and spatial distribution of cluster members. This holds even in the case of the modeling based on solutions of the spherically symmetric Jeans equation. 
Finally we compare all methods, using Bayesian inference, in order to determine which model best describes the cluster, and quantify the presence of dark matter. 

Our paper is organized as follows: in Section 2 we describe the data sets we are using, while in Section 3 we describe the creation of the synthetic brightness profile, the  likelihood method, the model selection using Bayesian evidence, how we test for rotation, and the dynamical modeling of the cluster. In Section 4 we present our results,  and our conclusions in Section 5.

\section{Data}
NGC 6809 is a Galactic globular cluster that lies in the constellation of Sagittarius. It has a roughly  spherically symmetric shape, and is at a distance $d\sim 5400$pc from the Sun \citep{1996AJ....112.1487H}. The full spatial extent of the cluster reaches out to $\sim 30$pc and the  coordinates of the cluster centre are (RA, Dec):  $(x_c,y_c)=(294.99879^{\circ},-30.96475^{\circ})$ as described in    the \cite{2000A&AS..143....9W} catalogue. 

In our analysis, we combined data sets from two independent sources. For the spatial distribution of the cluster and kinematics we used published values from  LA10; 
in their work the authors obtained spectra using observations performed with the AAOmega spectrograph on the Anglo-Australian Telescope (AAT). The position information for their targets was obtained from the Two Micron All Sky Survey (2MASS) Point Source Catalogue \citep{2006AJ....131.1163S}. 
The cluster membership of NGC 6809 stars was determined using the following criteria: (i) equivalent width of calcium triplet lines $w_{\lambda}$, (ii) surface gravity ($\log g$), (iii) 	line-of-sight velocity $v_{los}$  and (iv)  uncalibrated metallicity  ([m/H]). Targets that satisfied all the above criteria were characterized as cluster members. Out of 7462 targets in total, finally 728 stars were characterized as NGC 6809 cluster members. These we used for our dynamical analysis, i.e. we did not perform any additional selection of our own. 

The surface brightness distribution is more difficult as
no publicly available  data covering the range of radii of the kinematic data exists. 
Hence we create a synthetic surface brightness profile (hereafter SBP; see section \ref{DLI_synth_bright_prof}). This distribution is based on the published fit values from   \cite{1995AJ....109..218T}, where the authors have   combined data sets from various sources. They
used different techniques in order to create  a catalogue as accurate as possible for the surface brightness profiles in the $V$ band for 125 Galactic globular clusters. 
  CCD surface brightnesses make up the bulk of their data. Other techniques used include: digitized photographic and photoelectric photometry and photographic star counts in outer regions of star clusters. Their basis set is the Berkeley Globular Cluster Survey 
     \citep{1984ApJ...277L..49D,1986ApJ...305L..61D,1986IAUS..118..281D}, augmented by various contributions   \citep{1960AJ.....65..581K, 1979AJ.....84..505D, 1984PASP...96..198K} and others. 
For the SBPs \cite{1995AJ....109..218T} give reference values for a best fitting King profile:
\begin{equation}\label{DLI_ref_SB}
J(r)=J_0 \left(
\frac{1}{\sqrt{1+(r/r_c)^2}}-
\frac{1}{\sqrt{1+(r_t/r_c)^2}}
\right)^2
\end{equation}
($r_c$, $r_t$ are the core and tidal radius respectively) as well as quoted uncertainties for the defining parameters of the model \citep[][hereafter TR933; these currently appear to represent the most accurate 
surface brightness distribution for NGC 6809]{1993ASPC...50..347T}.  

\section[]{Dynamical Modeling}

\subsection{Synthetic Brightness Profile}\label{DLI_synth_bright_prof}
Here we  describe how we use the published reference values from TR93 in order to construct our SBP.
For the creation of a probability distribution for the brightness values  
we followed the following procedure: 
\begin{enumerate} 
\item Well within the radial region that corresponds in observational brightness estimates from TR93, we chose  35 radial positions $R_j$ in a uniform interval. Our choice of radial region was so as to avoid errors due to extrapolation from the reference King profile.  Based on the adopted distance of NGC 6809, we use values $R\in [0.26,  20.80]$pc.
\item For each position $R_j$, there corresponds a brightness value $J$ that depends on the defining parameters $\mu_V(0), r_c, c=\log_{10}(r_t/r_c)$ of the King brightness profile (Equation \ref{DLI_ref_SB}). For each given position $R_j$ we created a sample of $10^5$ random brightness values $J$ based on the reference values of the defining parameters and their quoted uncertainties $\delta\mu_V,\delta r_c,\delta c$ from TR93. 
\item Each $J$ value of the sample is created from Equation 1, for a  random value of the defining parameters $\mu_V(0), r_c, c=\log_{10}(r_t/r_c)$. These random values are drawn     from a Gaussian probability distribution  with mean $\mu$ equal to the best fitting reference value, and variance $\sigma$ equal to the uncertainty of the corresponding value.
\item 
From the sample of these $J$ values, we construct a normalized histogram that is equivalent to a probability distribution function $P_j(J)$. The index $j$ corresponds to the given position $R_j$.
\end{enumerate}

From the sample of all these $J$ values, for all positions $R_j$, we construct the synthetic surface brightness profile from which we can draw random data sets and associated errors.   With this approach we avoid any bias  from assuming a reference King profile. Note that  the distribution of values 
of the full brightness profile will be reflected in the uncertainty and distribution of parameter estimates for each model. Also, as it will become apparent in the next section, the brightness profile mainly constrains the mass-to-light ratio, and not the model parameters.

\subsection{Likelihood Analysis}
In this section we describe the construction of our likelihood function based on the data we use. We also give a short discussion on the motivation behind our formulation.
In the following we will be using standard  Bayesian approaches to model fitting and model selection; the reader is directed to standard texts such as \cite{sivia2006data} and \cite{2010blda.book.....G}  but we reproduce the details here for completeness. 

Let $\theta$ represent the vector of parameters needed to fully describe a given assumed physical model. In the framework of the Bayesian interpretation, we are interested in the posterior probability distribution of these parameters, taking into account our full data set $D=\{D_B, D_K \}$:
\begin{equation}
P(\theta | D)\propto P(\theta) \mathcal{L}(D|\theta)
\end{equation}
where $D_B$ is the brightness data and $D_K$ our kinematic data. $P(\theta)$ represents the probability of our prior range for the set of variables, and 
$\mathcal{L}(D|\theta)$ our likelihood model. For $P(\theta)$ we consider a uniform prior range 
 for all parameters, hence: 
 \begin{equation}
P(\theta)=\prod_{k=1}^{N_{\text{params}}} \frac{1}{\Delta \theta_{ k }}. 
 \end{equation}
 $N_{\text{params}}$ represents the total number of parameters and $\Delta \theta_{ k }$ the range of possible values for parameter 
$k$. 
Our likelihood model must take into account both the brightness and kinematics data. Since these two datasets are mutually independent it follows: 
\begin{equation}\label{DLI_4}
\mathcal{L}(D|\theta)=\mathcal{L}(D_B|\theta) \mathcal{L}(D_K|\theta)
\equiv \mathcal{L}_B \cdot \mathcal{L}_K
\end{equation}
and
\begin{align}\label{DLI_5}
\mathcal{L}_B &= 
\prod_{{ j }=1}^{N_{\text{rad}}} P_{  j   }(\Upsilon^{-1}\Sigma(R_{ j   }) )
\\ \label{DLI_6}
\mathcal{L}_K &= \prod_{i=1}^{N_{\star}} f(R_i,v_i).
\end{align}
$N_{\star}$ is the total number of stars,  $N_{\text{rad}}$ the total number of radial positions for our synthetic brightness data and $v_i$ the observed values of line-of-sight velocity ($v_{los}$). $P_{j}(J)$ is the probability of the value of 
brightness $J=\Upsilon^{-1}\Sigma(R_{ j })$ in bin $j$, as estimated by the normalized histogram of synthetic brightness values. 
The function $f$ is an  approximation  to the exact projected distribution function convolved with a Gaussian error $\delta v$ in the velocity. 
It is in principle feasible to obtain the projected distribution function $f(R,v_{los})$ for each one of the considered models (even numerically). 

However, for simplicity and consistency with the spherically symmetric Jeans equation solver\footnote{ 
In this case we can not find an analytic or numeric form for $f$}, we will use the following approximate form:
\begin{equation}\label{DLI_projected_PDF}
f(R_i,v_i)\approx
\frac{\Sigma(R_i)}{M_{\text{tot}}}
\frac{
\exp {\left[
 -\frac{(v_i- \langle v_{\text{los}} \rangle )^2}{2 (\sigma_{\text{los}}^2(R_i)
 +(\delta v_i)^2)}
\right]}
}{\sqrt{2\pi (\sigma_{\text{los}}^2(R_i) + (\delta v_i)^2)}}.
\end{equation}
$\Sigma(R)$ is the  projected mass density, $M_{\text{tot}}$ the total mass of the system, effectively a normalization constant.    $\langle v_{\text{los}} \rangle$ is the systemic velocity of the cluster, $v_i$ and $\delta v_i$ the observed star velocity and its error. $\sigma_{\text{los}}(R)$ is the  theoretical form of the 	line-of-sight velocity dispersion for the model under consideration.   
Combining Equations  \ref{DLI_4}, \ref{DLI_5}, \ref{DLI_6} and \ref{DLI_projected_PDF} we get: 
\begin{equation}\label{DLI_likelihood}
\mathcal{L}= 
\prod_{i=1}^{N_{\star}}
\frac{\Sigma(R_i)}{M_{\text{tot}}}
\frac{
\exp {\left(
 -\frac{(v_i- \langle v_{\text{los}} \rangle )^2}{2 (\sigma_{\text{los}}^2
 +(\delta v_i)^2)}
\right)}
}{\sqrt{2\pi (\sigma_{\text{los}}^2 + (\delta v_i)^2)}}
\prod_{j=1}^{N_{\text{rad}}}
P_j\left(\frac{\Sigma(R_j)}{\Upsilon} \right)
\end{equation}
where it should be clear that the index $i$ refers to each individual cluster member, while index $j$ refers to the position of each radial bin.

In order to estimate the highest likelihood values of the parameters $\theta$ we employed a Markov Chain Monte Carlo (MCMC) algorithm, namely a ``stretch move'' as described in \cite{GoodmanWeare}. This method has the advantage of exploring the parameter space more effectively, thus avoiding problems where
parameters are constrained around local maxima of the likelihood function. Our MCMC walks were run for sufficient autocorrelation time, so as to ensure that the distributions of parameters were stabilized around certain values. 

We would like to stress the importance of using the complete projected probability distribution, exact or approximate, for the estimation of the model parameters. 
In the projected PDF lies information for both the spatial extent of the system, and the values of velocities. Out of few thousand stars from LA10, a subsample of these is characterized as cluster members. Despite the fact that this subset is not the complete cluster, it obeys the same normalized spatial distribution. Just like any random subsample of a normalized distribution should. 
Looking at our likelihood (Equation \ref{DLI_likelihood}) 
the first product  alone can constrain the set of parameters of the  model dynamically, without making any use of the SB information.
In the opposite case, where we would have used a likelihood of the form: 
\begin{equation}\label{DLI_likelihood_bad}
\mathcal{L}= 
\prod_{i=1}^{N_{\star}}
\frac{
\exp {\left(
 -\frac{(v_i- \langle v_{\text{los}} \rangle )^2}{2 (\sigma_{\text{los}}^2
 +(\delta v_i)^2)}
\right)}
}{\sqrt{2\pi (\sigma_{\text{los}}^2 + (\delta v_i)^2)}}
\prod_{j=1}^{N_{\text{rad}}}
P_j\left(\frac{\Sigma(R_j)}{\Upsilon} \right)
\end{equation}
thus omitting the spatial distribution information $\frac{\Sigma(R_i)}{M_{\text{tot}}}$ we would have performed a bias overestimating the importance of velocity values. This would be due to the higher number of terms in the left product  compared to the number of brightness data. The second product of Equation  \ref{DLI_likelihood} constrains mainly the mass-to-light ratio, having a small effect on the marginalized distributions of the rest of the model parameters.

Furthermore in tests we performed with synthetic data we observed that the MCMC walks were converging faster when we were using the full projected PDF (Equation 
\ref{DLI_projected_PDF}). In this case  the marginalized model parameter distributions had also smaller variance.

\subsection{Bayesian Model Selection}
In our present description, we use Bayesian model selection \citep{2010blda.book.....G} and Nested Sampling (\citealt{Skilling}, hereafter JS04), a method for estimating the Bayesian evidence for a given likelihood model. For completeness, we give a short introduction to these methods. 

\begin{figure}
\centering
\includegraphics[width=\columnwidth]{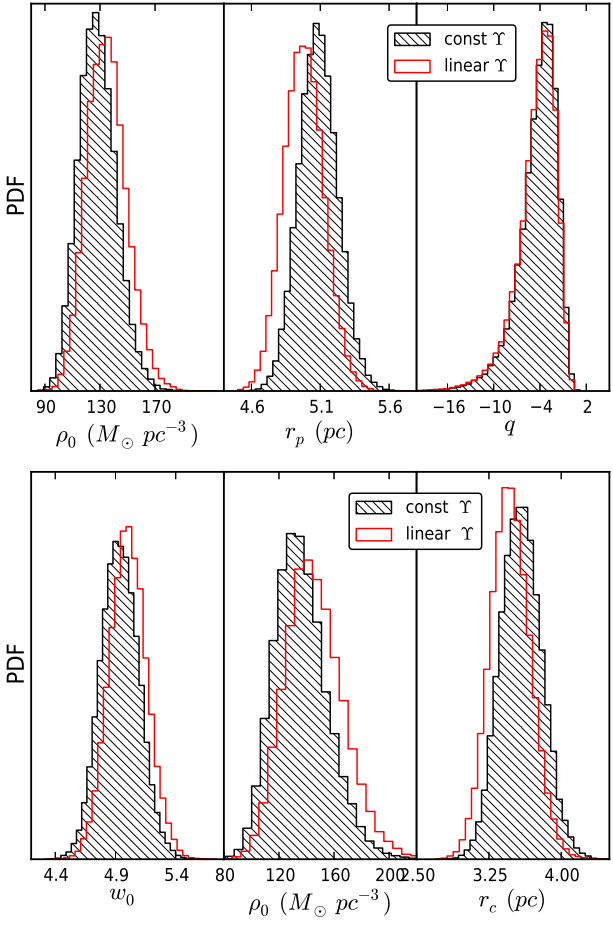}
\caption{Marginalized probability density distributions for the parameters of the Plummer (top) and King (bottom) models for both constant and linear mass-to-light ratio $\Upsilon(r)$. Shaded region corresponds to constant mass-to-light ratio. For both cases, i.e. $\Upsilon=$const or $\Upsilon=$ linear, similar results are obtained.}
\label{DLI_ModelsPDFparams}
\end{figure}

\begin{figure}
\centering
\includegraphics[width=\columnwidth]{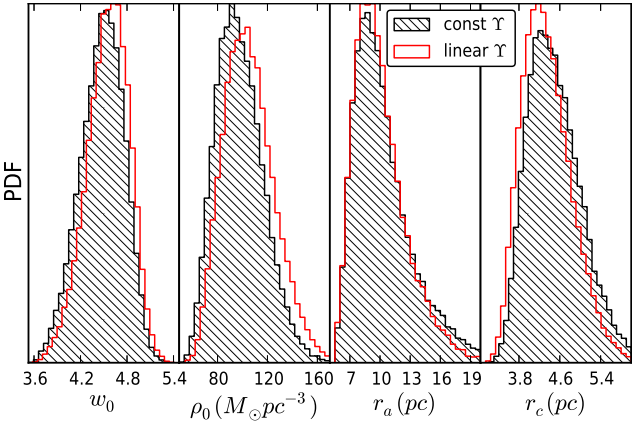}
\caption{Same as Fig. \ref{DLI_ModelsPDFparams} for Michie model.}
\label{DLI_Michie_PDFparams}
\end{figure}

Let $M_{\alpha }$ represent each of the models used in our analysis (e.g. $M_1\equiv$ Plummer with constant $\Upsilon$ etc). Furthermore let $I=M_1+\cdots + M_n$ represent our hypothesis, that at least one of the models is correct. Summation indicates logical ``or''. Let   $\theta$ represent the total number of parameters for each model and $D$ our data set. According to Bayes theorem the probability  of the model parameters $\theta$ given the data set of values is:
\begin{equation}
p(\theta | D,M_{  \alpha   },I)= \frac{p(\theta|M_{ \alpha},I) \mathcal{L}(D| \theta, M_{ \alpha    },I)}{p(D|M_{\alpha},I)}.
\end{equation}
$p(\theta|M_{ \alpha },I)$ is the prior information on the parameters,  $\mathcal{L}(D| \theta, M_{ \alpha  },I)$ 
is the likelihood as defined in Equation  \ref{DLI_likelihood}  and $p(D| M_{ \alpha },I)$ is the normalization constant for the model 
$M_{ \alpha }$ under consideration. 
This constant plays an important role for model selection. 
Marginalizing over all parameters, for the set of competing 
hypotheses, the probability of a model given  the data is:
\begin{equation}
p(M_{ \alpha } |D,I) = \frac{p(M_{ \alpha }|I) p(D|M_{\alpha},I)}{p(D|I)}
\end{equation}
Our level of ignorance of model choice suggests that $p(M_{\alpha }|I)=p(M_{\gamma}|I)$ for any $\alpha, \gamma $ combination (all models are equiprobable). Hence the relative ratio of probabilities of two models is: 
\begin{equation}
\frac{p(M_{ \alpha }   |D,I)}{p(M_{ \gamma}|D,I)} 
= \frac{p(M_{\alpha }|I) p(D|M_{ \alpha    },I)}{p(M_{ \gamma    }|I) p(D|M_{ \alpha    },I)}=
\frac{ p(D|M_{ \alpha    },I)}{p(D|M_{  \gamma    },I)}=O_{{  \alpha   }{   \gamma    }}
\end{equation}
$O_{{  \alpha   }{ \gamma}}$ is defined as the odds ratio, and it quantifies the comparison of two competing models for the description of observables.  
$p(D|I)$ is the normalization constant that does not participate in our calculations each time we compute the relative ratio of two models.  

Nested Sampling, introduced by JS04,  is an algorithm for the estimation of the normalization parameter 
$p(D|M_{  \alpha  },I)$. 
Following his terminology, the evidence $Z_{ \alpha  }$ of model $M_{\alpha}$ is given by:
\begin{equation} \label{Skilling_evidence}
Z_{  \alpha    }= p(D|M_{ \alpha },I) = \int p(\theta|M_{ \alpha    },I) \mathcal{L}(D| \theta, M_{\alpha    },I) d\theta 
\end{equation}
and corresponds to the normalization constant $p(D|M_{ \alpha   },I)$. 
Making use of the prior mass $dX=p(\theta|M_{ \alpha },I) d\theta$, an effective parameter transformation from $\dim (\theta) = n$ to $\dim(X) =1$, the above integral is simplified: 
\begin{equation}
Z_{ \alpha  }=\int_{0}^1 \tilde{\mathcal{L}}(D|X) dX
\end{equation}
In order to estimate this quantity and perform model selection we use MultiNest \citep{2008MNRAS.384..449F,2009MNRAS.398.1601F}.  This algorithm is designed for effective calculation of Bayesian evidence based on Skilling's algorithm. It gives consistent results even in the case of multimodal likelihood functions.

Assuming we have a complete set of models, the following system of equations 
allows us to have an estimate for the probability $p(M_{\alpha }|D,I)$ for each individual model:
\begin{align*}
\sum_{{\alpha}=1}^{N_{\text{M}}} p(M_{  \alpha  }|D,I) &=1 \\
\frac{p(M_1|D,I)}{p(M_2|D,I)} &= O_{12}\\
& \vdots \\ 
\frac{p(M_1|D,I)}{p(M_{N_{\text{M}}}|D,I)} &= O_{1N_{\text{M}}}\\
\end{align*}
where $N_{\text{M}}$ is the total number of models considered. 

\begin{figure*}
\centering
\includegraphics[width=\textwidth]{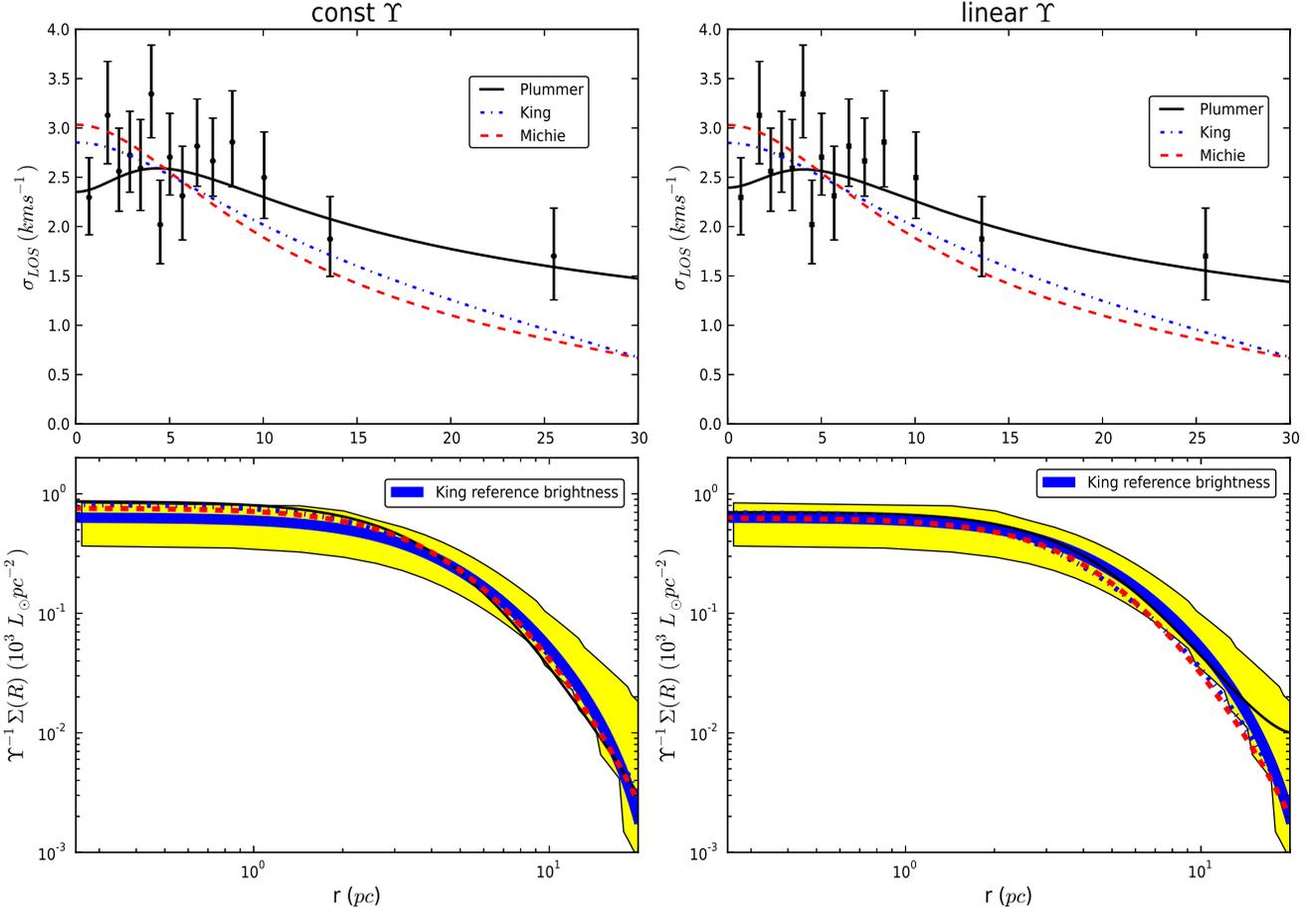}
 \caption{Highest likelihood fitting models for  constant (left panels) and  linear (right panels) mass-to-light ratios. 
 The King reference brightness corresponds to the best values quoted by Trager et al. (1995). 
 The yellow colored regions corresponds to 1  $\sigma$ variation  as estimated by the distribution of brightness values in the corresponding points.
In the top  panels the $\sigma_{los}$ data points correspond to estimates from Lane et al. (2010). 
We should stress however that in our parameter estimates, we took account of the full kinematic profile, i.e. all  observed 
line-of-sight velocities, and not a binning scheme. The drawn lines correspond to the various $\sigma_{los}(r)$ 
functions for the highest likelihood fit of the various PDF models.}
\label{DLI_PDFModels_bright_sigmaLOS}
\end{figure*}

\subsection{Cluster Rotation}
It is important to consider whether NGC 6809 undergoes significant rotation and, if so, do we have to take this into account in our analysis.  We consider the following modification of the likelihood (equation  \ref{DLI_likelihood})
\begin{equation} \label{DLI_rotation}
\mathbf{v}_{los} \to \mathbf{v}_{los} + \bomega \times \mathbf{R}
\end{equation}
where $\boldsymbol{\omega}$ is a vector representing angular rotation. The axis of rotation  lies on the $(\xi, \eta)$ plane of tangent coordinates and is defined by the angle $\phi$ between the positive $\bxi$ direction and
 $\bomega=\omega \; \hat{\mathbf{r}}$. That is, we add two additional parameters in the likelihood model, $\omega, \phi$.

The criterion for the need to drop the assumption of spherical symmetry for the cluster, is going to be the relative values of the actual potential, $\Phi(r)$, compared to the centrifugal potential, $-\frac{1}{2}|\bomega \times \mathbf{r}|^2$, throughout the extent of the cluster. In the case where rotation is significant, the Collisionless Boltzmann Equation needs to be  modified  \citep{2008gady.book.....B} according to 
\begin{equation*}
\frac{\partial f}{\partial t} + \mathbf{v}\cdot \nabla f
- \left(2 (\bomega \times \mathbf{v}) + \nabla \Phi_{\text{eff}} \right)\cdot \frac{\partial f}{\partial \mathbf{v}} = 0
\end{equation*}
where $\Phi_{\text{eff}}=\Phi - \frac{1}{2} |\bomega \times \mathbf{r}|^2$ and $f$ the distribution function of the system. 

\subsection{Models with a known Distribution Function}

Here we give a short introduction of  the mathematical formulation for the dynamical models we use.
We consider three distinct models that describe with fair accuracy the brightness profiles and dynamics of globular clusters, namely  \citet{1911MNRAS..71..460P}\footnote{see also \citet{1987MNRAS.224...13D}},  \cite{1966AJ.....71...64K} and \cite{1963MNRAS.125..127M} models. 
For each model, we consider two different functional forms for the mass-to-light-ratio; these are a constant $\Upsilon(r)=$const and a linear\footnote{It should be clear that $\Upsilon_a$ is the slope of the mass-to-light ratio, and $\Upsilon_b$ the intercept.}  
 model with $\Upsilon(r)=\Upsilon_a\; r + \Upsilon_b$. The latter of these two cases is not meant to be physical. This is because a negative slope $\Upsilon_a <0$ would yield $\Upsilon \leq 0$ for some $r>0$. However, a positive slope $\Upsilon_a$ would indicate that the cluster is embedded in a dark matter halo.  On the other hand, a large negative slope $\Upsilon_a$ in combination with a high central value of mass-to-light ratio $\Upsilon(r=0) = \Upsilon_b$ could be suggestive of massive dark remnants in the core.   

\subsubsection{Plummer Model}
A general anisotropic Plummer model is defined from a probability distribution function that depends only on the energy per unit mass of the system $f(\mathcal{E})$. It is fully described by the set of parameters $(\rho_0, r_p, q)$. 
$\rho_0$ is the core density at the centre ($r=0$), $r_p$ is the characteristic Plummer radius, and $q$ the anisotropy parameter. Here we only state the equations we used for consistency with units. For a full analytic description of Plummer models the reader should consult    \cite{1987MNRAS.224...13D}. 

Demanding the potential $\Phi_P(r)$ to be consistent with the mass density through 
Poisson's equation $\nabla^2 \Phi_P=4\pi G \rho(r)$ yields: 
\begin{align}
\Phi_P(r)&= -\frac{4 \pi  r_p^2 G \rho_0}{3 \sqrt{1+(r/r_p)^2}}\\
\rho(r)&=\frac{\rho_0}{(1+r^2/r_p^2)^{5/2}}\\
\sigma_{\text{los}}^2(R)&=\frac{\pi ^2 G \rho_0 r_p^3 \left(12 r_p^2+(12-5 q) R^2\right)}{32 (6-q) \left(r_p^2+R^2\right){}^{3/2}}
\end{align}
The quantities that are finally tested against observational data are the projected mass density 
\[
\Sigma(R)=2 \int_{R}^{\infty} \frac{\rho(r) r }{\sqrt{r^2-R^2}} dr
\] 
and the line-of-sight velocity dispersion $\sigma_{\text{los}}(R)$.

\subsubsection{King and Michie Models}
King and Michie models are derived from a distribution function that depends on energy $\mathcal{E}$ and, in the case of Michie, also on angular momentum \footnote{It is to be understood that when we mention energy or angular momentum this is per unit mass.} $L$. King is an isotropic model, i.e. the anisotropy parameter\footnote{Here we consider that in a spherical coordinate system $(r,\theta,\phi)$, the tangential velocity dispersion is defined as $\sigma_t^2=\sigma_{\theta}^2+\sigma_{\phi}^2$.}   is $\beta=1-\sigma_t^2/(2\sigma_r^2)=0$, therefore more restrictive.  
Michie allows a more general class of trajectories yet not completely arbitrary.
The analysis for both probability distribution functions is very similar hence we  follow a unified description for both.

King and Michie models  are derived from the following PDFs:
\begin{align}\label{M55_KingDF}
f_K(\mathcal{E})&=
\begin{cases}
\frac{f_{0K} }{ (2\pi \sigma_K^2)^{3/2} } 
\left(
e^{-\mathcal{E}/\sigma_K^2} -1
\right) & \mathcal{E}<0 \\
0 & \mathcal{E} \geq 0
\end{cases}\\
f_M(\mathcal{E},L)&=
\begin{cases}
\frac{f_{0M}}{(2\pi \sigma_M^2)^{3/2}} 
e^{-\frac{L^2}{2\sigma_M r_a^2}}
\left(
e^{-\mathcal{E}/\sigma_M^2} -1
\right) & \mathcal{E}<0 \\
0 & \mathcal{E} \geq 0
\end{cases}
\end{align}
where $f_{0J}, \sigma_J$,  $J \in \{K,M\}$  and $r_a$ are parameters to be determined from Bayesian  likelihood methods. 

Let $r_t$ denote the tidal radius of the system, i.e. a position beyond  which the mass density and all physical quantities of the system vanish. If $\Phi(r)$ is the potential, by making use of an arbitrary additive constant to it's definition, we may define as a new potential the difference: $\Psi=\Phi(r)-\Phi (r_t)$; now $\Psi$ vanishes 
at the tidal radius. Furthermore, in order to simplify our calculations, we introduce the transformation\footnote{Index $J\in \{K,M\}$, $K$ for King, and $M$ for Michie model.}:  
$w_J=-\Psi(r)/ \sigma_J^2$, respectively for King and Michie models.
Then: 
\begin{equation}
\mathcal{E}_J=\frac{v_r^2+v_t^2}{2}-2\sigma^2_J w_J(r)
\end{equation}
The following functions can be calculated analytically for both models with the use of  Computer Algebra Systems (e.g. Maxima, Mathematica, Maple), as functions of radius $r$ and ``potential'' $w(r)$ :
\begin{align}
\rho(r,w)&=4 \pi \int_{v_r=0}^{\sqrt{2\sigma^2 w}} 
\int_{v_t=0}^{\sqrt{2\sigma^2 w-v_r^2}} f(\mathcal{E},L)  v_t dv_t dv_r
\\
\sigma_{r}^2(r,w)& = \frac{4\pi}{\rho}\int_{v_r=0}^{\sqrt{2\sigma^2 w}} 
\int_{v_t=0}^{\sqrt{2\sigma^2 w-v_r^2}} v_r^2 f(\mathcal{E},L) v_t dv_t dv_r \\
\sigma_{t}^2(r,w)& = \frac{4\pi}{\rho}\int_{v_r=0}^{\sqrt{2\sigma^2 w}} 
\int_{v_t=0}^{\sqrt{2\sigma^2 w-v_r^2}}  f(\mathcal{E},L) v_t^3 dv_t dv_r .
\end{align}
$v_r$ is the radial component of the velocity in spherical coordinates $(v_r,v_{\theta},v_{\phi})$ and $v_t^2=v_{\theta}^2+v_{\phi}^2$. 
In order to facilitate comparisons to the observed data, we convert these functions into a projected mass density and 
a line-of-sight velocity dispersion:

\begin{align}
\Sigma(R)&= 2\int_{r=R}^{r_t} \frac{\rho(r,w) r}{\sqrt{r^2-R^2}}dr\\
\sigma_{\text{los}}^2(R)&=\frac{1}{\Sigma(R)}
\int_{r=R}^{r_t} \frac{
 \rho \; (2  \sigma_r^2 (r^2-R^2) 
+  \sigma_t^2 R^2 )
}{r \sqrt{r^2-R^2}} dr
\end{align}

\begin{figure}
\centering
\includegraphics[width=\columnwidth]{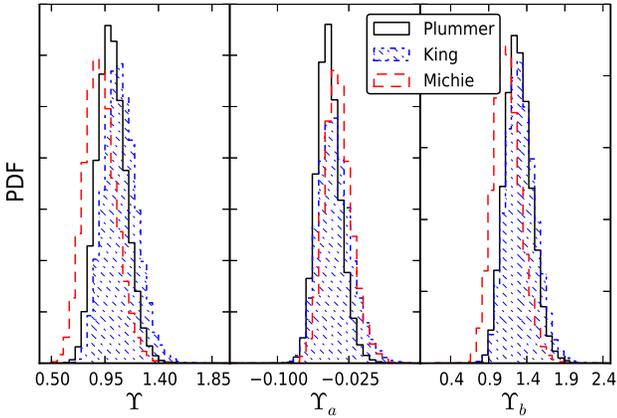}
 \caption{The histogram on the left is the marginalized  $\Upsilon$ distribution for the case of constant mass-to-light ratio for the various PDF models. The other two correspond to the marginalized  $\Upsilon_a$ and  $\Upsilon_b$ distributions for the case of linear mass-to-light ratio $\Upsilon(r)=\Upsilon_a\; r+\Upsilon_b$, again for the corresponding PDF based models.}
\label{DLI_PDFModels_combinedY}
\end{figure}

A  model is fully described once we assign values to it's defining parameters  and know the functional form of the ``potential'' $w(r)$. 
The latter is achieved by solving Poisson's equation numerically. To do this, we require two additional assumptions at $r=0$: an initial value for the potential $w_0$ and the equilibrium condition $\frac{d w }{dr} \bigr|_{r=0}=0$.

Instead of $(f_{0J},\sigma_J)$ 
 it is very convenient to use  the mass core density $\rho_0$ and the King core radius
$r_c$ defined by:
\begin{align*}
\rho_0& = \rho\left(r,w(r)\right) \bigr|_{r=0}, & 
r_{cJ} &= \left(
\frac{9 \sigma_J^2}{4 \pi G \rho_0}
\right)^{1/2}
\end{align*}
Then for the full description of a King or a Michie model  we use the following set of parameters: 
\begin{align*}
(w_0,\rho_0,r_c) \qquad &\text{King}\\
(w_0,\rho_0,r_c, r_a) \qquad &\text{Michie}
\end{align*}
Using the transformed potential $w$, the Poisson equation is most conveniently written:
\begin{align}
 \nabla^2 w(r)&= - \frac{9}{ r_c^2} \tilde{\rho}(r,w), &
 \text{where}\quad 
 \tilde{\rho}(r,w) & = \frac{\rho(r,w)}{\rho_0}
\end{align}
The steps followed for a full evaluation of a 
 model are the following: 
\begin{enumerate}
\item Assign initial values to parameters, $(w_0,\rho_0,r_c)$ for King   and  $(w_0,\rho_0,r_c,r_a)$ for Michie models.
\item Subject to the initial conditions $w(r=0)=w_0$ and  $\frac{dw}{dr}|_{r=0}=0$ 
solve Poisson's equation numerically, 
thus obtain $w(r)$.
\item Quantities $\rho(r), \sigma_r(r), \sigma_t(r)$ are fully determined upon knowledge of $w(r)$. Hence also $\Sigma(R)$ and $\sigma_{\text{los}}^2(R)$ that will be used for comparison with observations. 
\end{enumerate}

\begin{figure}
\centering
\includegraphics[width=\columnwidth]{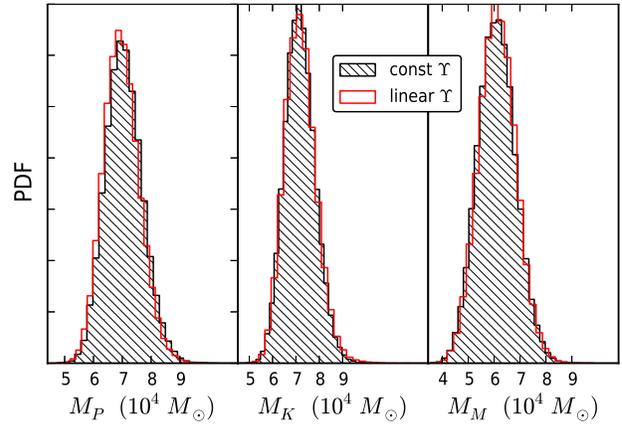}
\caption{Mass estimates for the various PDF based models. From left to right: Plummer, King and Michie mass distributions. Shaded region corresponds to $\Upsilon=$ const.}
\label{DLI_ModelsPDFMass}
\end{figure}

\subsection{Modeling based on solutions of the \\* Spherically  Symmetric Jeans  Equation}
In this approach, we consider that the stellar system, and a possible dark matter component are gravitationally bound, self supported and spherically symmetric. Furthermore we make the assumption that the system consists of a single stellar component, 
which is sufficiently accurate for our purposes here \citep{1981QJRAS..22..227K}. Let $\rho_{\star}$ and  $\rho_{\bullet}$  be the mass densities respectively of stellar and dark matter. We describe the stellar component by a Plummer mass density, while for the dark matter one, we use a Navarro, Frenk and White (hereafter NFW) profile \citep{1996ApJ...462..563N}: 
\[
\rho_{\bullet}(r)= \frac{r_s^3 \rho_{0\bullet}}{r (r_s^2+r^2)^2}. 
\]
Then the  spherically symmetric Jeans equation reads: 
\begin{equation}\label{DLI_Jeans_1}
\frac{1}{\rho} \frac{d (\rho \sigma_{r}^2)}{dr} + \frac{2 \beta }{r} \sigma_r^2 = -\frac{GM(r)}{r^2}
\end{equation}
where it must be clear that we make the assumption of linear addition of mass densities $\rho=\rho_{\star}+\rho_{\bullet}$. Then the kinematic quantities $\sigma_r^2$ and $\beta=1-\sigma_t^2/(2\sigma_r^2)$ describe the possibly compound structure as a whole\footnote{$\beta>0$ corresponds to radial anisotropic profile, while $\beta<0$ to a tangential one.}.

\begin{figure}
\centering
\includegraphics[width=\columnwidth]{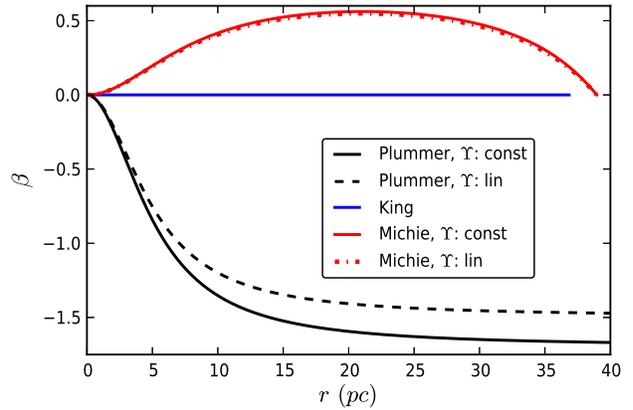}
\caption{Highest likelihood $\beta$ anisotropy profiles for the various PDF models. The King and Michie models are truncated to the value of their corresponding tidal radius $r_t$ for the best fitting models.}
\label{DLI_betaPDFs}
\end{figure}

Let $\lambda_{\beta}(r)$ be the integrating factor in Equation \ref{DLI_Jeans_1} i.e. 
$\lambda_{\beta}(r) = \exp (2 \int (\beta/r) dr)$. Then $\sigma_r^2(r)$   is given by: 
\begin{equation}
\sigma_r^2 (r)= \frac{1}{\lambda_{\beta} \rho(r)} \int_r^{\infty} 
\rho(s) \lambda_{\beta}(s) \frac{G M(s)}{s^2} ds 
\end{equation}
where we have assumed that $\sigma_r \to 0$ as $r\to \infty$.
The connection with observables is performed through the 	line-of-sight velocity dispersion (resulting from both, stellar and dark matter distributions)
\begin{equation}
\sigma_{los}^2(R) = \frac{2}{\Sigma(R)} \int_{R}^{\infty} 
\left( 1-\beta(r) \frac{R^2}{r^2} \right) \frac{r \; \rho \; \sigma_r^2}{\sqrt{r^2-R^2}} dr
\end{equation}
and the stellar projected mass density $\Sigma_{\star}$. The likelihood in this approach is written: 
\begin{equation}
\mathcal{L}=\prod_{i=1}^{N_{\star}} \frac{\Sigma_{\star}(R)}{M^{\star}_{\text{tot}}}
\frac{\exp (- \frac{(v_i-\langle V_{los}\rangle)^2 }{2 (\sigma_{los}^2+(\delta v_i)^2)})}
{\sqrt{2 \pi (\sigma_{los}^2+(\delta v_i)^2)}} 
\prod_{j=1}^{N_{\text{bins}}} P_j(\Sigma_{\star}(R_j))
\end{equation}
where again $P_j(\Sigma_{\star}(R_j))$ is the probability of the value 
$\Sigma_{\star}(R_j)$ in the brightness  histogram in position $R_j$. 

\begin{figure}
\centering
\includegraphics[width=\columnwidth]{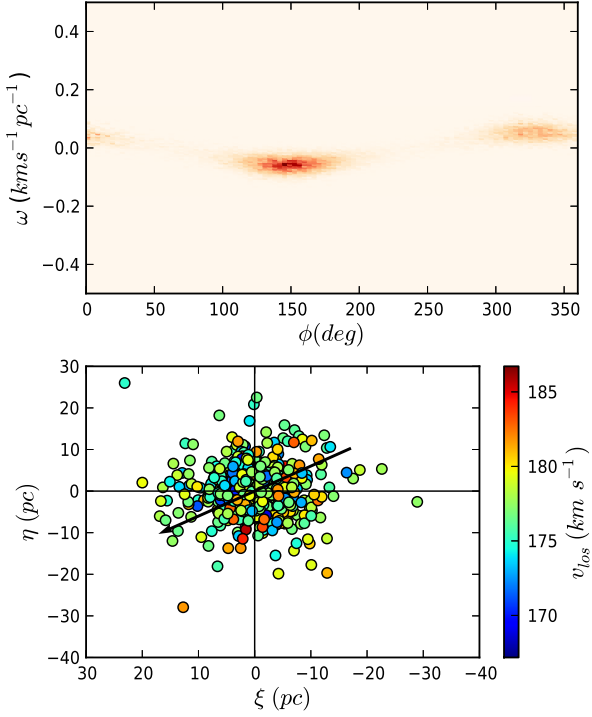}
 \caption{Top panel: Density plot of MCMC chains for $\omega$, $\phi$ parameters that define the rotation vector $\bomega$.  Bottom panel: Positions of cluster members of M55 on the tangent plane and their corresponding line-of-sight velocity $v_{los}$. The vector corresponds to the $\bomega$ angular rotation in arbitrary scale. }
\label{DLI_Rotation_vector}
\end{figure}

Since we do not know the form of the $\beta$ anisotropy profile, we assume two distinct cases, a constant and a Plummer-like functional form: 
\begin{align}
\beta_1 &\equiv \text{const}\\
\beta_2(r) &= \frac{q}{2} \frac{r^2}{r_{\beta}^2+r^2}. 
\end{align}
$q$ and $r_{\beta}$ are free parameters, different from the corresponding Plummer stellar profile,  to be determined by our statistical analysis. $q/2$ is the limiting value we expect to find as $r \to \infty$. 
The $\beta_1$ approximation defines a severe restriction on the system (making it less probable to exist), nevertheless it can reveal the existence of a dark matter component. On the other hand, guided by the  Osipkov-Merritt form $\beta(r)=r^2/(r^2_{\beta}+r^2)$ which is valid for a wide range of systems (see \citealt{2008gady.book.....B}), we expect to have a better fit through the use of $\beta_2(r)$. 

\section{Results}

\subsection{Rotation}
With the assumption of a Plummer model with constant or linear mass-to-light ratio 
$\Upsilon$, we estimated the parameters $(\omega, \phi)$. We quote the following values that are in accordance with results from LA10: 
\begin{align*} 
\omega &= -0.05 \pm 0.025 \;\rm{ (km\; s^{-1}\; pc^{-1})}\\
\phi &= 148.97 \pm 29.80 \;\rm{(deg)}
\end{align*}
In Fig. \ref{DLI_Rotation_vector} we visualize our results where  we plot the density map of the MCMC chains for parameters $\omega$ and angle $\phi$ as well as the direction of the rotation vector $\bomega$. We emphasise that in Fig. \ref{DLI_Rotation_vector} the magnitude of $\bomega$ is magnified in order to be visible and does not account for the real magnitude  of the vector which is small.    The extent of our cluster is up to $\backsim 30\; \rm{pc}$. Since the majority of cluster members are within $\sim 20\; \rm{pc}$ we expect that rotation has no significant role in the gravitational acceleration ($\frac{1}{2}(\omega r)^2 < 1.0$ within 20 pc distance). Indeed, comparing this value with the potential values $\Psi = - \sigma^2 w(r)$ for either a King or Michie model, it is significantly smaller. Hence we conclude there is no need to model the cluster with potential altered by rotation. Furthermore, due to axial symmetry the rotation effect will not change our estimate of the systemic velocity of the cluster, or any of the model parameters. In tests we performed, we  recovered the same distributions of values for all the parameters, irrespective of whether we modified the likelihood according to Equation \ref{DLI_rotation} or not. 

\subsection{Model Comparison using Bayesian Evidence}

\begin{figure}
\centering
\includegraphics[width=\columnwidth]{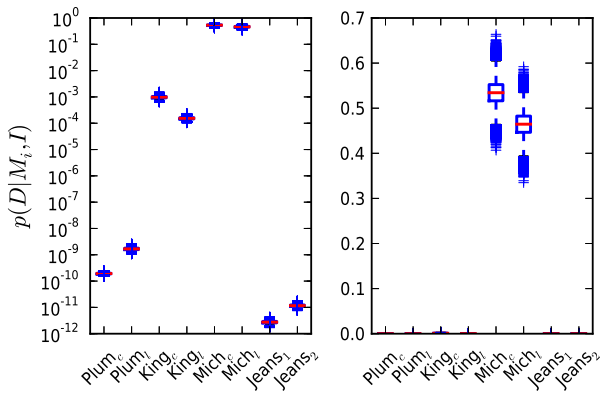}
\caption{
Relative probabilities of competing models. The central red mark is the median. The edges of each box are the lower hinge (defined as the 25th percentile) and the upper hinge (75th percentile). Whiskers extend to the most extreme data points. We plot in normal (right figure) and $\log$ (left figure) scale. 
}
\label{DLI_EvidenceProbs}
\end{figure}

\begin{figure}
\centering
\includegraphics[width=\columnwidth]{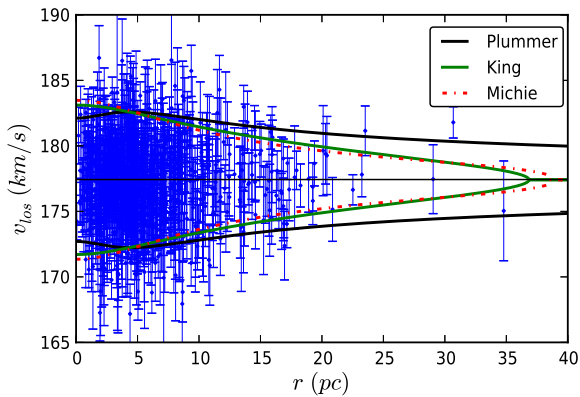}
 \caption{Line of sight velocity values $v_{los}$ for various distances $r$ from the centre of the cluster. Overplotted are the $\pm 2\, \sigma_{los}(r)$ for the highest likelihood fitting models, centreed at the systemic velocity of the cluster $v_{los}=177.43^{+0.17}_{-0.09}$. In contrast with Fig. \ref{DLI_PDFModels_bright_sigmaLOS} where the Plummer model appears to give the best fit for the $\sigma_{los}$ plot, it is seen that  King and Michie models have a tidal radius close to the more distant cluster members, while Plummer extrapolates to much further distances.
  }
\label{DLI_Vlos_vs_sigmaLOS_fits}
\end{figure}

\begin{table}
\caption{Bayesian evidence  }
\label{DLI_Evidence}
\begin{center}
\begin{tabular}{@{}  l   c   c  c }
\hline 
Model  &  $\log(Z) \pm \delta \log(Z)$ & $p \pm \delta p$
 \\ \hline \hline
Plummer, $\Upsilon:$ const    & $  -5848.25 \pm 0.11$ & $ (1.92 \pm 0.10) \times   10^{-10}$\\ \hline
Plummer, $\Upsilon:$  linear  &  $  -5846.09      \pm 0.12 $& $ (16.70 \pm 1.60)\times   10^{-10}$ \\ \hline
King, $\Upsilon:$ const   & $-5832.81 \pm 0.11$& $ (9.75 \pm 1.00)\times   10^{-4}$  \\ \hline
King, $\Upsilon:$  linear    & $-5834.66 \pm 0.09$&   $ (1.52 \pm 0.10)\times   10^{-4}$ \\ \hline
Michie, $\Upsilon:$ const     & $-5826.51 \pm 0.10$& $0.53 \pm 0.03$ \\ \hline
Michie, $\Upsilon:$ linear &  $ -5826.65 \pm  0.11$& $0.47 \pm 0.03$\\ \hline
Jeans, $\beta_1$&    $-5852.50 \pm  0.12 $&  $ (2.75 \pm 0.30)\times   10^{-12}$ \\ \hline
Jeans, $\beta_2$& $-5851.06  \pm 0.11$& $ (1.16 \pm 0.13)\times   10^{-11}$\\ \hline
\end{tabular}
\end{center}
\medskip

Left column: name of model. Middle: Bayesian evidence with uncertainty as estimated from MultiNest. Right column: relative probability of model, with quoted error, based on the assumption that at least one of the models is correct.  
\end{table} 

Based on the values estimated by MultiNest we report the following (see Table  \ref{DLI_Evidence} and Fig. \ref{DLI_EvidenceProbs}): The most favored models are the Michie Models. The least probables are the ones based on the solution of the spherically symmetric Jeans  equation. In general all of the PDF based models appear to have significantly higher probability than the Jeans based models. However it should be clear that this is related to our choice of given mass models and anisotropy parameters, not to Jeans equations methods in general. 
For the case of the Michie model, there can be no clear distinction between the constant or linear mass-to-light ratio $\Upsilon$. This follows from the fact that the two models have probability values within error bars in the same interval. In Table \ref{DLI_Evidence} we give the estimated values of the evidence $\log Z$ as well as the corresponding probability value of each model. In Fig. \ref{DLI_EvidenceProbs} we visualize these results by creating the box plot of the corresponding probabilities, both in normal (left panel) and log scale (right panel). The edges of each box are the lower hinge (corresponds to the 25th percentile) and the upper hinge (corresponds to 75th percentile). Whiskers correspond to the most extreme values of the probabilities. 

Comparing   $\sigma_{los}$ fit among the various PDF models in Fig. \ref{DLI_PDFModels_bright_sigmaLOS}, the Plummer model seems to give a better fit. However if we plot all of the 	line-of-sight values of velocities versus distance from the cluster centre (Fig. \ref{DLI_Vlos_vs_sigmaLOS_fits}) we note two things: First the large uncertainty of errors in measurement in velocities dominates in the outer regions where there exist few data points, thus resulting in a large total $\sigma_{los}$. Second, the Plummer model extends much further than our last data points as seen in Fig. \ref{DLI_Vlos_vs_sigmaLOS_fits}; this is in direct contrast with King or Michie models. Thus the Plummer model extrapolates out to possible structure beyond the last data points, something that Nested Sampling penalizes. 

\subsection{Models with a known Distribution Function}

\begin{table*}
\caption{PDF based Models }
\label{DLI_PDFparams}
\centering
\begin{tabular}{@{}  l| l l | l l l l l l l }
 \hline 
 Model & $w_0$ & $\rho_0 (M_{\odot}pc^{-3})$ &$r_p (pc)$ & $r_c (pc)$ & $q$ & $r_a (pc)$
 & $\Upsilon_a$ & $\Upsilon_b$ & Mass $(10^4\, M_{\odot})$\\ \hline \hline
 \multirow{2}{*}{Plummer } &  & 
 $126.87_{-16.03}^{+13.36} $ 
&$5.10_{-0.18}^{+0.10}$ & & $-2.86_{-3.80}^{+ 0.84}$   &  
&  & $0.99_{- 0.13}^{+0.13}$ & $ 7.05^{+0.50}_{-0.75} $ \\[6pt] 
  &  
&$131.67_{- 12.88}^{+ 15.46}$
& $4.97^{+0.12}_{-0.17}$    & & 
$-3.23^{+1.27}_{-3.39}$ & & 
$-0.05^{+0.01}_{-0.01}$ & 
$1.24^{+0.19}_{-0.17}$ & $6.83 ^{+0.71}_{-0.60}$ \\ \hline
 \multirow{2}{*}{King }  & 
$4.91^{+0.17}_{-0.17}$&  
$132.54^{+19.91}_{-19.91}$&   & 
$3.55^{+0.20}_{-0.23}$& &  & 
 & $1.04^{+0.14}_{-0.12}$ & $ 7.10 ^{+0.58}_{-0.70}$\\[6pt] 
 & 
$4.99^{+0.14}_{-0.17}$& 
$139.75^{+23.36}_{-19.46}$&  & 
$3.43^{+0.23}_{-0.19}$& &  & 
$-0.04^{+0.02}_{-0.02}$ & 
$1.30^{+0.19}_{-0.19}$ & $7.10 ^{+0.65}_{-0.81}$\\ \hline
\multirow{2}{*}{Michie} & 
$4.48^{+0.23}_{-0.35}$&  
$91.52^{+24.15}_{-14.50}$ & &
$4.24^{+0.74}_{-0.34}$ &   &
$8.76^{+4.80}_{-1.50}$  & 
 & $0.90^{+0.14}_{-0.14}$ & $6.10^{+0.51}_{-0.88}$ \\[6pt] 
 & 
$4.66^{+0.11}_{-0.42}$& 
$101.64^{+21.58}_{-19.18}$ & &
$4.20^{+0.60}_{-0.41}$ & &
$8.46^{+5.00}_{-1.20}$ & 
$-0.04^{+0.02}_{-0.02}$ & 
$1.09^{+0.21}_{-0.18}$ & $ 6.00^{+0.59}_{-0.73}$\\ \hline
\end{tabular}
\medskip

Estimates of parameters for PDF based models within  $1\sigma$ interval. For the mass of the Plummer model we took the limiting case where $r \to \infty$. For King and Michie models we estimated the mass contained up to the tidal radius of the system. We introduced $\Upsilon(r) = \Upsilon_a \; r+ \Upsilon_b$, then for each model the first row corresponds to a constant mass-to-light ratio ($\Upsilon_a=0$), while  the second to a linear $\Upsilon(r)$.
\end{table*} 

\begin{figure}
\centering
\includegraphics[width=\columnwidth]{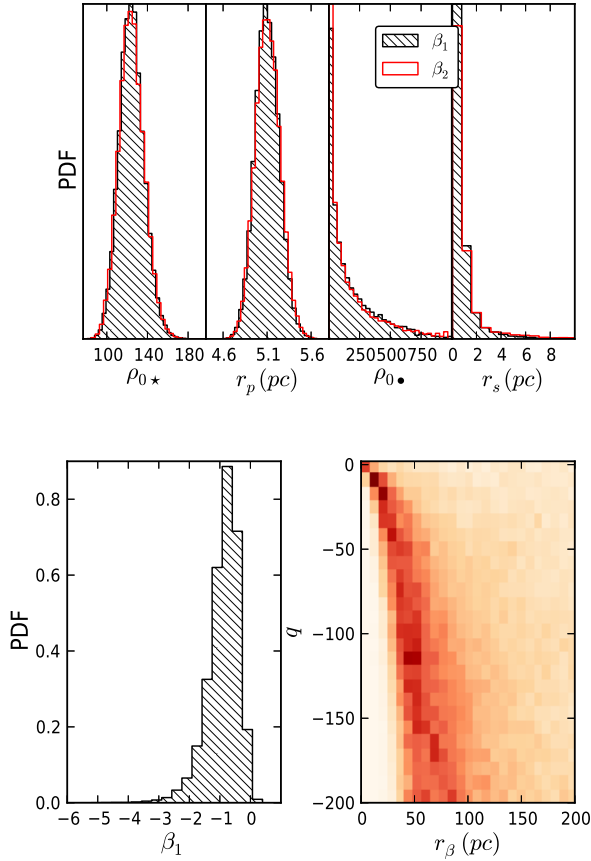}
\caption{Top panel: Marginalized distributions of Jeans model parameters. 
From left to right: Stellar core density, characteristic Plummer radius, dark matter core density and NFW characteristic radius.
Bottom left panel: Marginalized distribution of  $\beta_1$ parameter. Bottom right: Density map created from the points in the MCMC walk for the case of $\beta_2$. It is evident that  parameter  $q$ demonstrates a degeneracy, while $r_\beta$ has a clear constrained distribution of values.}
\label{DLI_ModelsJeansparams}
\end{figure}

All of the PDF based models suggest that there is no significant amount of dark matter in NGC 6809. This result is valid for both a constant and a linear mass-to-light ratio. As expected the parameter marginalized distributions (Fig. \ref{DLI_ModelsPDFparams} and \ref{DLI_Michie_PDFparams})  are the same  irrespective of the choice of $\Upsilon(r)$.
This verifies that they are constrained from the LA10 kinematic set and not the King reference brightness profile. Furthermore, for the case of the linear $\Upsilon(r)$, the slope is small, thus comparable to the case with constant $\Upsilon$. In Fig. \ref{DLI_PDFModels_combinedY} we plot the marginalized distributions of the $\Upsilon(r)$ various models. 

The negative slope rejects any hypothesis for the cluster to be embedded in a DM halo. 
If there existed a DM halo surrounding the cluster, the mass-to-light ratio should increase with distance from the cluster centre, i.e. we should have a possitive slope $\Upsilon_a$ for the linear $\Upsilon(r)$.
Furthermore, the small value of mass-to-light ratio at $r=0$, i.e. $\Upsilon_b$, rejects the possibility of a massive DM remnant in the core.
For the best fitting Michie models, the constant mass-to-light ratio at $r=0$ is $\Upsilon = 0.90^{+0.14}_{-0.14}$, while the linear $\Upsilon_b = 1.09^{+0.21}_{-0.18}$ within 1$\sigma$ confidence interval. Both values suggest there is no significant dark matter core remnant. Even in a $95\%$ condifence interval, the values again for Michie model are for constant  $\Upsilon = 0.90^{+0.27}_{-0.25}$  and for the linear case  at $r=0$  $\Upsilon_b = 1.09 ^{+ 0.41}_{-0.33}$.   
The maximum likelihood values, for which we performed the fits lie well within the region of quoted uncertainty in Table \ref{DLI_PDFparams}. 

Our mass estimates are  close to the order of magnitude as estimated by LA10 and \cite{2009A&A...500..785K}.  Yet we find, within  the error bars,  a factor of 2 smaller value for the modal value of total mass, according to our best fitting model (Michie).  Specifically LA10 quotes $1.4\pm 0.5 \times 10^5 M_{\odot}$ while we find $6.00^{+0.59}_{-0.73} \times 10^4 M_{\odot}$, i.e. approximately half the modal value. This is due to the fact that LA10 used an isotropic Plummer model for the fit, and also because their parameter estimates took into account only the kinematic 	
line-of-sight $v_{los}$, not the brightness. Furthermore, LA10 used a binning scheme that has the disadvantage of small number of data points, and also larger uncertainty due to error propagation. We adopted the same isotropic Plummer profile and performed a calculation with our method and recovered the same mass value and Plummer radius $r_p$ as LA10. That is, the restriction to isotropy in the Plummer model reveals a slightly higher mass estimate and a significantly altered scale radius $r_p$. 	In Fig. \ref{DLI_ModelsPDFMass} we give the marginalized distributions of Mass for each model. Finally in Fig. \ref{DLI_betaPDFs} we plot the $\beta$ anisotropy profiles of the models that correspond to the highest likelihood values. Despite the fact that the Plummer model can have radial or tangential $\beta$ anisotropy\footnote{$\beta>0$ corresponds to radial anisotropic profile, while $\beta<0$ to tangential.}, it demonstrates a tangentially anisotropic profile, in direct contrast with the best fitting Michie model, which is radially anisotropic. 

Comparing our work to \cite{2005ApJS..161..304M}, we find a significantly smaller $\Upsilon$; it is smaller than all their dynamical quoted values (ranging from a King fit $3.23^{+1.42}_{-1.18}$ to $2.83^{+1.25}_{-1.02}$ for a power-law model).  There are several reasons for this:
\begin{enumerate}
\item We use both isotropic (King) and anisotropic (Plummer, Michie, Jeans models) distributions while they use only isotropic ones. 
\item We have a greater kinematic sample (they used $\sim$20 values from \cite{1991AJ....102.1026P} in comparison with 728 cluster members in the present work)
\item The uncertainties, as quoted from  \cite{1993ASPC...50..347T}, for our brightness model allow for a smaller mass-to-light ratio if the central brightness values are higher than the best fitting King model. However, our approach gives a unique modal value for each brightness point. That is any deviation from the brightness modal values reflects the need for the model to more accurately describe the 	
line-of-sight velocity dispersion. 
\item Our estimates of  mass-to-light ratio are very close to those from 
\cite{1991AJ....102.1026P} within the lowest error bound. The authors model the cluster based on a Michie model, but do not perform comparison of different models. Moreover they have  different brightness profile and kinematic samples, and a different  likelihood method for estimating parameters.
\item  $\Upsilon$ is estimated mainly from the brightness reference profile and its quoted uncertainties. Thus the marginalized parameter distributions will also reflect the uncertainties in the SBP.
\end{enumerate}
Finally  LA10 quotes  $\Upsilon = 2.0^{+0.9}_{-0.8}$ which in its lower bound is close to our predicted value. It is notable that we also find smaller value for $\Upsilon$ from the predicted value of canonical mass-to-light ratio \cite{2009A&A...500..785K}, the authors quote  $2.03 \pm 0.02$. 

Recently \cite{2012ApJ...755..156S} performed mass estimates for the same cluster, based on the same kinematic data set and using a substantially  different method. Their estimates of mass and $\Upsilon$ are a little higher than ours,  $\Upsilon = 1.6 \pm 0.2$ and $M_{\text{dyn}} \approx 1.6 \pm 0.1 \times 10^5 M_{\odot}$. The authors used a subsample of stars from \cite{2010MNRAS.401.2521L} for their kinematic data, namely stars satisfying $d<2 r_h$ in order to avoid overestimation of velocity dispersion due to tidal heating in the outskirts of the cluster. We would like to note that our best fitting models, King or Michie, give a very small value for $\sigma_{los}$ in large distances, as seen in Fig. \ref{DLI_PDFModels_bright_sigmaLOS} and Fig. \ref{DLI_Vlos_vs_sigmaLOS_fits}, thus penalizing such an effect. 

\subsection{Modeling based on spherically  symmetric \\* Jeans Equation}
For both cases of distinct anisotropy $\beta$ parameters (e.g. Fig. \ref{DLI_ModelsJeansparams}) the marginalized PDF of the dark matter  parameters  tends to zero values ($r_s$ value is related to the form of  spherically symmetric Jeans equation).  
In order to compare the masses of the stellar and dark matter population, we followed the following procedure; since the NFW dark matter profile does not posses a finite mass, we adopted a tidal radius far greater than the physical extent of the cluster, i.e. $r_t=100$ pc. For the whole range of acceptable parameters from the MCMC walk we estimated the mass of the DM and stellar  population up to the tidal radius $r_t$.  In a $95\%$ confidence interval the maximum value of dark matter mass in both cases does not exceed 150 $M_{\odot}$. The modal value for both cases of $\beta$ anisotropy assumptions is $\sim 2.5 M_{\odot}$ something clearly negligible. We consider that it is much more likely this small amount of matter to result from observational constrains rather than a separate DM component, or that the assumption of linear addition of mass densities is not a viable one. 

For the two distinct cases of $\beta$ functional forms, the $\beta_1$=const is clearly least favored from Nested Sampling (see Table \ref{DLI_Evidence}). 
For the case of the $\beta_2$ functional form, there is a degeneracy in values of the $q$ parameter. Actually $q \to -\infty$, suggesting that the model favors circular trajectories. 

As mentioned previously, the Jeans models are the least favored from Bayesian evidence  for the specific choices of anisotropy parameter $\beta$, thus we rule out the existence of any separate dark matter component. We should point out however that this analysis is different from the one followed in \cite{2013MNRAS.428.3648I}. The authors in their work did not assume a specific anisotropy profile and used a very large number of parameters that made inevitable the calculation of Bayesian evidence.

\section{Conclusions}
We modeled NGC 6809 using the most complete kinematic sample to date  \citep{2010MNRAS.401.2521L}. We employed a variety of models in order to explore the possibility of a dark matter component.  Namely models with a known distribution function and models that are evaluated through the use of the  spherically symmetric Jeans equation. 
 For our modeling we used the complete projected PDF, thus all of the kinematic data, and not a binning scheme in order to avoid error propagation and having to use a larger data set. This results, due to a higher number of data points, in more constrained parameter estimates. We note that our results will be improved with future precision photometry of the large-scale stellar population in NGC 6809. 
Additionally, we tested the cluster for rotation, using a Bayesian  likelihood method finding a small value of angular velocity $\omega = -0.05 \pm 0.025\; \rm{km\; s^{-1}\; pc^{-1}}$, thus concluding it is not necessary to model the cluster with a rotating potential. 

For the comparison of several models we used Nested Sampling, a method that estimates the Bayesian evidence between competing hypotheses.
Model selection allows us to conclude, based on our hypothesis, if a DM component is present or not. 
 We report from our methods as the most probable by orders of magnitude the Michie models. The least favorable are the models based on the solution of the Jeans  spherically symmetric equation, that include a separate dark matter component.  The corresponding probabilities are for the Michie of the  order of $\sim 1.0$ while for the Jeans models of the order of $\sim 10^{-13}$. 
 
The negative slope $\Upsilon_a$ of the linear mass-to-light ratio  excludes the existence of a DM halo surrounding the cluster. The most favorable models from Bayesian inference give the smallest mass estimate and smallest dynamical mass-to-light ratio. For a Michie model with constant $\Upsilon$ we find $M_{\text{dyn}}=6.1^{+0.51}_{-0.88} \times 10^4 M_{\odot}$ 
 and $\Upsilon = 0.90^{+0.14}_{-0.14}\sim 1.0$ in 1$\sigma$ confidence interval.   In a $95\%$ confidence interval, the corresponding mass-to-light ratio is $\Upsilon =0.90^{+0.27}_{-0.25}$ thus suggesting there is no dark matter component in NGC 6809. 
 
 For a Michie  model with  linear $\Upsilon$  we find (1$\sigma$ confidence interval) $M_{\text{dyn}}=6.00^{+0.59}_{-0.73} \times 10^4 M_{\odot}$  and  mass-to-light ratio at the origin $(r=0)$ $\Upsilon_b = 1.09^{+0.21}_{-0.18}$. Again, at the $95\%$ confidence interval, this is  $\Upsilon_b = 1.09^{+0.41}_{-0.33}$, excluding a massive dark matter core remnant.
  
Our mass and dynamic mass-to-light ratio estimates are smaller than predictions from previous studies of NGC 6809, but are better  constrained.  \citet{2010MNRAS.401.2521L} quotes  $\Upsilon = 2.0^{+0.90}_{-0.80}$ while \citet{2005ApJS..161..304M} have estimates ranging from a King fit $\Upsilon = 3.23^{+1.42}_{-1.18}$ to $\Upsilon = 2.83^{+1.25}_{-1.02}$ for a power-law model. Both studies have a significantly higher modal value and a larger uncertainty than our estimates.

\section*{Acknowledgments}
F. I. Diakogiannis acknowledges the University of Sydney International Scholarship
for the support of his candidature. 
G. F. Lewis acknowledges support from ARC Discovery Project (DP110100678) and Future Fellowship (FT100100268). 
The authors would like to thank the anonymous referee for 
the useful comments and suggestions that helped us 
significantly improve our work. 
The authors would also like to thank Pascal Elahi, Magda Guglielmo, Richard Lane and Matthias Redlich for useful comments and discussion.

%
\bibliographystyle{mn2e} 
\bibliography{DLI_paper.bib} 

\bsp

\label{lastpage}

\end{document}